\documentclass[%
superscriptaddress,
amsmath,amssymb,
aps,
prl,
floatfix,
twocolumn]{revtex4-2}

\usepackage{hyperref}
\usepackage{amsmath}
\usepackage{amssymb}
\hypersetup{colorlinks=true,
    linkcolor=blue,
    filecolor=blue, 
    citecolor=blue,
    urlcolor=blue,
}
\usepackage{graphicx}
\usepackage{braket}
\usepackage{bm}

\usepackage{dcolumn}
\usepackage{enumitem}
\usepackage{mathtools}
\usepackage{color}
\usepackage{xcolor}
\usepackage{verbatim}
\usepackage{float}
\usepackage{bbold}

\newcommand{\abs}[1]{\lvert #1 \rvert} 
\newcommand{\ev}[1]{\langle #1 \rangle} 
\newcommand{\lv}[0]{\mathcal{L}} 
\newcommand{\td}[1]{\tilde{#1}} 
\newcommand{\hSig}[0]{\hat{\sigma}} 
\newcommand{\eps}[0]{\epsilon}

\begin{document}

\newcommand{\iitm}{\affiliation{Department of Physics, Indian Institute of Technology Madras, Chennai, India, 600036.}}
\newcommand{\iisc}{\affiliation{Department of Instrumentation and Applied Physics, Indian Institute of Science, Bangalore, India, 560012.}}
\newcommand{\optimas}{\affiliation{Physics Department and Research Center OPTIMAS, University of Kaiserslautern-Landau, D-67663, Kaiserslautern, Germany}}

\title{Quantum Synchronization and Dissipative Quantum Sensing}

\author{Gaurav M. Vaidya}
\iitm
\author{Simon B. J\"ager}
\optimas
\author{Athreya Shankar}
\email[Correspondence: ]{athreyas@iisc.ac.in}
\iisc

\date{\today}

\begin{abstract}
We study the phenomenon of quantum synchronization from the viewpoint of quantum metrology. By interpreting quantum self-sustained oscillators as dissipative quantum sensors, we develop a framework to characterize several aspects of quantum synchronization. We show that the quantum Fisher information (QFI) serves as a system-agnostic measure of quantum synchronization that also carries a clear operational meaning, viz., it quantifies the precision with which the amplitude of a weak synchronizing drive can be measured. We extend our analysis to study many-body oscillators subjected to multiple drives. We show how the QFI matrix can be used to determine the optimal drive that maximizes quantum synchronization, and also to quantitatively differentiate the synchronization responses induced by different drives. Our work highlights multiple connections between quantum synchronization and quantum metrology, paving a route towards finding quantum technological applications of quantum synchronization. 
\end{abstract}

\maketitle

\emph{Introduction.}--- Classical synchronization is a long studied and ubiquitous phenomenon in nature and forms the basis of numerous everyday applications. More recently, several studies have explored synchronization in quantum systems and its relation to quantum features such as coherence, interference and entanglement~\cite{lee2013quantum,walter2014quantum,lee2014entanglement,xu2014synchronization,hush2015PRA,zhu2015NJP,lorch2017quantum,nigg2018observing,roulet2018synchronizing,roulet2018quantum,sonar2018squeezing,koppenhofer2019PRA,koppenhofer2020quantum,jaseem2020quantum,laskar2020observation,parra2020synchronization,krithika2022observation,buca2022SciPost,tan2022quantum,zhang2023quantum,shen2023quantum,murtadho2023cooperation,nadolny2023macroscopic,solanki2023symmetries,shen2023quantum}. One notion of quantum synchronization (QS) explores the development of coherences in a dissipatively stabilized quantum system with no \emph{a priori} phase preference, when it is subjected to a phase-symmetry breaking perturbation~\cite{hush2015PRA,roulet2018synchronizing,roulet2018quantum,koppenhofer2019PRA,jaseem2020quantum,laskar2020observation,tan2022quantum,krithika2022observation}. A central challenge here is to devise metrics to quantify the synchronization response of quantum systems~\cite{ameri2015PRA,roulet2018synchronizing,jaseem2020generalized}. At the same time, the question of how QS may be broadly relevant for quantum technology applications remains largely unanswered, in part because existing quantifiers of QS are either not universally applicable or do not admit a clear operational interpretation in the context of specific quantum information processing tasks. 

In this Letter, we demonstrate that the toolbox of quantum Fisher information (QFI)~\cite{pezze2018RMP,liu2020QuantumFisher} provides a systematic way to quantify and analyze several aspects of QS in a system-agnostic manner. Conversely, quantum synchronizing systems can naturally be interpreted as dissipative quantum sensors~\cite{reiter2017NatComm, dutta2019PRL, xie2020PRApp} tasked with measuring the amplitude of a symmetry-breaking perturbation. We illustrate the central ideas interfacing QS and dissipative quantum sensing using the paradigmatic example of a quantum van der Pol oscillator~\cite{lee2013quantum}. Subsequently, we extend our approach using the QFI matrix (QFIM) to study many-body oscillators with multiple drives. Using a minimal example of a two-qubit oscillator~\cite{vaidya2024exploring}, we show how our framework allows to optimize the combination of applied drives to maximize QS, and also enables to quantify differences in the synchronization responses induced by the different drives. More broadly, our work reinforces the utility of the QFIM as a diagnostic tool in diverse contexts beyond its original setting of quantum metrology (QM)~\cite{liu2020QuantumFisher}.

\emph{Framework of quantum synchronization.}---We consider an open quantum system whose intrinsic dynamics is governed by a Lindblad master equation $\partial_t \hat{\rho} = \lv_0 \hat{\rho}$. The Liouvillian $\lv_0$ has the general form ($\hbar=1$)
\begin{eqnarray}
    \lv_0\hat{\rho} = -i[\hat{H}_0,\hat{\rho}] + \sum_k \mathcal{D}[\hat{O}_{g,k}]\hat{\rho} +
    \sum_k \mathcal{D}[\hat{O}_{d,k}]\hat{\rho},
\end{eqnarray}
where $\hat{H}_0$ is the free system Hamiltonian, and $\hat{O}_{g,k},\hat{O}_{d,k}, \; k=1,2,\ldots$ are one or more jump operators that respectively set up gain and damping channels via Lindblad dissipators of the form $\mathcal{D}[\hat{O}]\hat{\rho}=\hat{O}\hat{\rho}\hat{O}^\dag-\hat{O}^\dag\hat{O}\hat{\rho}/2-\hat{\rho}\hat{O}^\dag\hat{O}/2$. In order to constitute a valid quantum synchronizing system, $\lv_0$ and the steady state $\hat{\rho}_0$ must have an underlying $U(1)$ symmetry that reflects the absence of a preferred phase. This symmetry will be evident in the model systems we introduce below; for a formal discussion, see the Supplementary Material (SM)~\cite{supp}. 

We consider the synchronization of this system to a weak $U(1)$-symmetry breaking perturbation that changes $\lv_0\to\lv_0+\eps\lv_1$. Accordingly, the steady state of the system is changed to leading order as $\hat{\rho}_0\rightarrow\hat{\rho}_0+\eps \partial_\eps\hat{\rho}$, where the differential change is given by  
\begin{eqnarray}
    \partial_\eps\hat{\rho} = -\lv_0^{-1}\lv_1\hat{\rho}_0.
    \label{eqn:drho}
\end{eqnarray}
Here, $\lv_0^{-1}$ is computed excluding its null space, and we have assumed that $\lv_0$ has a unique steady state $\hat{\rho}_0$ and that all of its other eigenmodes are damped. In our examples below, the perturbations are of the form $\eps\lv_1\hat{\rho} = -i\eps[\hat{H}_1,\hat{\rho}]$, where $\hat{H}_1$ is a (dimensionless) $U(1)$-symmetry breaking Hamiltonian. The above framework includes, but is not restricted to, notions of QS that consider systems whose unperturbed steady states are diagonal in the eigenbasis of $\hat{H}_0$, e.g., as in Refs.~\cite{roulet2018synchronizing,jaseem2020quantum,jaseem2020generalized,koppenhofer2019PRA,krithika2022observation}. 

\emph{QFI measure for quantum synchronization.}--- We consider the above setting from a QM perspective. First, we note that in many QM settings, a state $\hat{\rho}$ is prepared as input to an interferometer and subsequently a signal to be sensed is encoded via unitary dynamics. This kind of interferometric approach, using intuitive choices for the unitary operator, has previously been employed to propose a QS  measure~\cite{shen2023fisher}. In contrast, we consider QS from a \emph{dissipative} quantum sensing scenario, where the signal to be sensed is directly the amplitude $\eps$ of the perturbation $\lv_1$, which is concurrently applied while the system is still subjected to $\lv_0$. We find this approach advantageous because it clearly distinguishes intrinsic dynamics from the synchronizing perturbation, doesn't rely on intuition to define the QS measure, and extends naturally to the case of multiple drives.

Writing the spectral decomposition $\hat{\rho}_0=\sum_k q_k \ket{k}\bra{k}$ for the unperturbed steady state, the QFI for sensing the amplitude $\eps$ from the change in the steady state is~\cite{helstrom1969quantum,holevo2011probabilistic,pezze2018RMP} 
\begin{eqnarray}
    \mathcal{F}[\lv_0,\lv_1] = \sum\limits_{\substack{k,k'\\q_k+q_{k'}>0}} \frac{2}{q_k+q_{k'}}\abs{\braket{k'|\partial_\eps \hat{\rho}|k}}^2,
    \label{eqn:qfi_1}
\end{eqnarray}
with $\partial_\eps \hat{\rho}$ given by Eq.~(\ref{eqn:drho})~\footnote{Typically, the QFI is written as a function of the initial state $\hat{\rho}$ prior to signal encoding, and the perturbation. However, assuming the steady state $\hat{\rho}_0$ is unique, Eq.~(\ref{eqn:qfi_1}) is independent of the initial state and instead depends only on the dynamics establishing $\hat{\rho}_0$. We emphasize this aspect by explicitly writing the $\lv_0$ dependence in Eq.~(\ref{eqn:qfi_1}), but henceforth suppress it for brevity.}. Equation~(\ref{eqn:qfi_1}) is our proposed QS measure, which also carries a clear operational meaning: The ultimate precision limit  on estimating $\eps$ using $K\gg 1$ independent copies of this system, optimized over all measurement strategies, is given by $(\Delta\eps)^2=1/(K\mathcal{F})$. In proposing Eq.~(\ref{eqn:qfi_1}) as a QS measure, we have assumed that $\lv_1$ only introduces coherences between $\ket{k},\ket{k'}$ belonging to different $U(1)$ symmetry sectors, which is true in many typical QS settings including the examples below (see SM~\cite{supp} for the general case).

Often, QS measures quantify the distance between the steady states in the presence and absence of a synchronizing drive, and hence explicitly depend on the drive amplitude $\eps$~\cite{roulet2018synchronizing,jaseem2020generalized}. Complementary to these measures, the QFI is a measure of the rate of change in the steady state with $\eps$, and is hence independent of $\eps$. Thus, it captures the small-signal response of the oscillator and is a direct indicator of the propensity of an oscillator to synchronize.

\begin{figure}
    \centering
    \includegraphics[width=0.7\columnwidth]{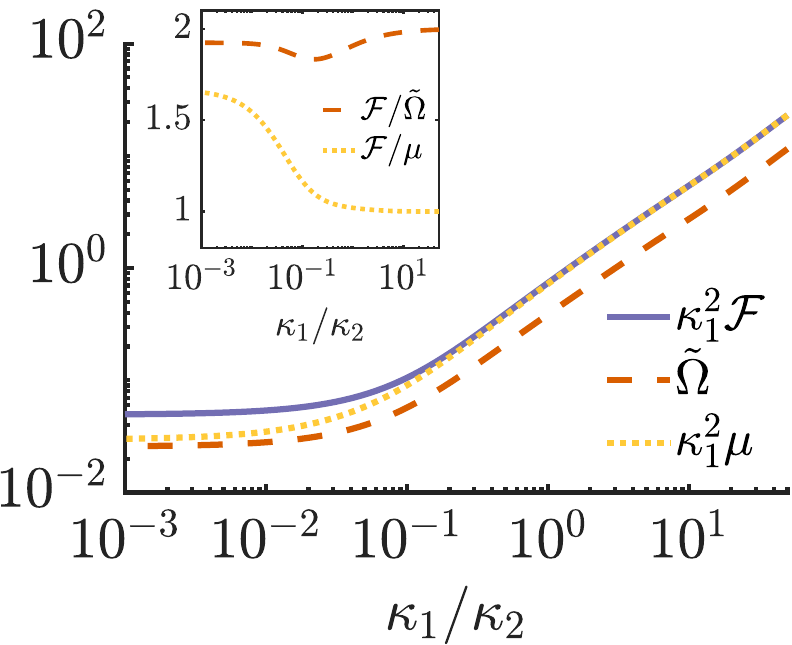}
    \caption{\textbf{Quantum synchronization in the quantum van der Pol oscillator.} The normalized QFI ($\kappa_1^2\mathcal{F}$) is compared against two other measures--- an entropy-based measured of QS ($\td{\Omega}$) and a metrology-based measure of the system response ($\kappa_1^2 \mu$)--- as the system is tuned from the quantum ($\kappa_1/\kappa_2\ll 1$) to the classical limit ($\kappa_1/\kappa_2\gg 1$), keeping $\kappa_1$ fixed. Inset: Ratios of the QFI to the other two measures, showing asymptotic proportionality in the classical limit ($\mathcal{F}/\td{\Omega}\to 2$, $\mathcal{F}/\mu \to 1$).}
    \label{fig:vdp_spin1}
\end{figure}

\emph{Example: Quantum van der Pol oscillator.}--- We consider a quantum van der Pol (vdP) oscillator with $\hat{H}_0=\omega\hat{a}^\dag\hat{a}$, $\hat{O}_{g}=\sqrt{\kappa_1}\hat{a}^\dag$, and $\hat{O}_d=\sqrt{\kappa_2}\hat{a}^2$. The $U(1)$ symmetry operator for this system is $\hat{U}(\phi)=e^{-i\phi\hat{a}^\dag\hat{a}}$.
As the symmetry breaking perturbation, we consider a resonant external drive, such that in the drive frame we have $\hat{H}_0=0$ and $\eps\hat{H}_1=\eps\hat{x}\equiv\eps(\hat{a}+\hat{a}^\dag)/2$. In Fig.~\ref{fig:vdp_spin1}, we plot $\kappa_1^2\mathcal{F}$ as $\kappa_1/\kappa_2$ is increased by fixing $\kappa_1$ and reducing $\kappa_2$. As $\kappa_1/\kappa_2$ becomes large, the limit cycle amplitude increases as $\ev{\hat{a}^\dag\hat{a}}^{1/2}\sim\sqrt{\kappa_1/\kappa_2}$ and the phase noise decreases since the phase diffusion constant is proportional to $ \kappa_2$~\cite{lee2013quantum}. As a result, stronger synchronization--- characterized by increasing $\mathcal{F}$--- is observed as $\kappa_1/\kappa_2$ increases. This behavior is consistent with the expectation that a self-sustained oscillator with large amplitude and low phase noise is highly sensitive to a weak, injection-locking resonant drive.

To demonstrate that $\mathcal{F}$ serves as a reliable QS measure, we compare it with a recently proposed information theoretic QS measure~\cite{jaseem2020generalized}. The latter is shown as the dashed line in Fig.~\ref{fig:vdp_spin1}(a) and is defined as $\Omega = S(\hat{\rho}) - S(\hat{\rho}_\mathrm{diag})$. Here, $S(\hat{\rho})$ is the von Neumann entropy for the steady state $\hat{\rho}$ under $\lv_0+\eps\lv_1$ and $\hat{\rho}_\mathrm{diag}$ is the reference `limit-cycle' state obtained by deleting all off-diagonal elements from $\hat{\rho}$. While $\mathcal{F}$ is independent of  $\eps$, $\Omega\propto \eps^2$ and hence we plot $\td{\Omega}=\Omega/\eps^2$ to enable comparison. The behavior of $\mathcal{F}$ is  qualitatively similar to $\td{\Omega}$ over the entire range of $\kappa_1/\kappa_2$. Furthermore, in the classical limit $\kappa_1/\kappa_2\gg 1$, $\mathcal{F}$ is asymptotically proportional to $\td{\Omega}$, as shown in the inset. In fact, we find that $\td{\Omega}\to \mathcal{F}/2$ quite generally, provided the steady-state populations and coherences vary smoothly across the ladder of Fock states, which is the case in the classical limit of this system~\cite{supp}. 

The above comparison suggests that, for this paradigmatic system, the QFI is on par with $\td{\Omega}$ as a QS measure. Moreover, and in contrast to $\td{\Omega}$, it also sets the ultimate precision for estimating the drive amplitude. In the classical limit, the value of $\mathcal{F}^{-1}$ agrees with the precision of intuitive estimation strategies based on measuring low-order moments of simple observables. The dotted line shows the quantity $\kappa_1^2\mu$, where $\mu=(\partial_\eps\ev{\hat{p}})^2/\mathrm{Var}(\hat{p})$, with $\hat{p}\equiv(\hat{a}-\hat{a}^\dag)/(2i)$ and $\mathrm{Var}(\hat{p}) = \ev{\hat{p}^2}-\ev{\hat{p}}^2$. In the method of moments approach to estimation~\cite{pezze2018RMP}, $\mu$ corresponds to the signal-to-noise ratio in estimating $\eps$ by measuring $\ev{\hat{p}}$ when the drive (along $\hat{x}$) is applied. While $\mu\leq\mathcal{F}$ for any $\kappa_1/\kappa_2$, the equality is saturated asymptotically for $\kappa_1/\kappa_2\gg 1$ (see inset), where we find that $\mu,\mathcal{F}\to 4/(9\kappa_1\kappa_2)$~\cite{supp}. 

In contrast, in the quantum limit $\kappa_1/\kappa_2\ll 1$, $\mathcal{F}$ and $\mu$ differ significantly, and estimation strategies based on classical intuition are suboptimal. The origin of this deviation can be quantitatively explained by accounting for the discrete quantum levels of the system. Analyzing the system using the lowest three Fock states, we find that 
$\mathcal{F}\to 4/(81\kappa_1^2)$, while $\mu\to 4/(135\kappa_1^2)$~\cite{supp}. 

The above example demonstrates how the QFI interfaces QS and QM: On the one hand, the behavior of $\mathcal{F}$ agrees with existing QS measures. On the other, by inherently optimizing over all measurement strategies, $\mathcal{F}$ captures the full sensitivity of the system to the synchronizing drive in all parameter regimes. Hence, it serves as an unambiguous QS measure that doesn't rely on intuition, while simultaneously carrying a clear operational meaning in the context of QM.

\emph{Synchronization under multiple drives.}--- In a many-body quantum system that undergoes synchronization, it is natural to consider the case of several external drives that are, e.g., applied on different constituents of the system. A systematic framework to explore QS in such a multiple-drive scenario has not been introduced so far. Here, we show how the QFI toolbox readily extends to this general setting.

We consider the simultaneous action of $M$ weak perturbations $\boldsymbol{\lv}_1=(\mathcal{L}_{1,1},\ldots,\lv_{1,M})$ with amplitudes $\boldsymbol{\eps}=(\eps_1,\ldots,\eps_M)$ that change $\lv_0\to\mathcal{L}_0+\boldsymbol{\eps}\cdot\boldsymbol{\lv}_1$. Then, the steady state changes in leading order according to $\hat{\rho}_0\rightarrow\hat{\rho}_0+\boldsymbol{\eps}\cdot \nabla_{\boldsymbol{\eps}}\hat{\rho}$, where $\nabla_{\boldsymbol{\eps}}\hat{\rho}$ is the generalization of Eq.~(\ref{eqn:drho}). An $M\times M$ QFI matrix (QFIM) $\boldsymbol{\mathcal{F}}_M$ can be constructed to analyze QS in this setting, with elements~\cite{liu2020QuantumFisher,helstrom1969quantum,holevo2011probabilistic} 
\begin{eqnarray}
    \mathcal{F}_{M,mn} =   \sum\limits_{\substack{k,k'\\q_k+q_{k'}>0}} \frac{2}{q_k+q_k'}\mathrm{Re}[\braket{k'|\partial_{\eps_m} \hat{\rho}|k} \braket{k|\partial_{\eps_n} \hat{\rho}|k'}].
    \label{eqn:qfim_1}
\end{eqnarray}
We discuss two applications of the QFIM to QS.

\emph{Drive optimization.}--- The $M$ simultaneous drives can be viewed as a single composite drive $\lv_{1,c}$ with strength $\eps_c$ that changes $\lv_0\to \lv_0 + \eps_c \boldsymbol{n}\cdot \boldsymbol{\lv}_1$. Here, $\boldsymbol{n}$ is a unit vector such that $\eps_j=\eps_c n_j$ are the individual drive amplitudes. The QS measure for the composite drive is just the QFI for sensing $\eps_c$, given by $\mathcal{F} = \boldsymbol{n}^T \boldsymbol{\mathcal{F}}_M \boldsymbol{n}$. We can immediately identify the optimum linear combination $\boldsymbol{n}_\mathrm{opt}$ of the $M$ drives that maximizes $\mathcal{F}$ as the eigenvector of $\boldsymbol{\mathcal{F}}_M$ corresponding to the largest eigenvalue. This observation extends the recent idea of using optimal unitary generators to encode parameters~\cite{reilly2023PRL} to the scenario of dissipative quantum sensing. Thus, in contrast to other QS measures, the QFI framework enables us to easily identify strategies to maximize QS in a given system.

\emph{Eigendrives and drive orthogonality.}---The QFI framework can be used to assess the distinguishability of the responses induced by two or more drives. To do so, we first note that for a given $\lv_0$ and $\boldsymbol{\lv}_1$, the eigenvectors of the QFIM can be understood as \emph{eigendrives} that drive mutually orthogonal modes of synchronization, in the following sense. We consider the parameter space spanned by $\boldsymbol{\eps}$, where each point is associated with the steady-state density matrix $\hat{\rho}_{\boldsymbol{\eps}}$. In this space, the QFIM $\boldsymbol{\mathcal{F}}_M$ defines a metric tensor at the origin $\boldsymbol{\eps}=\boldsymbol{0}$; each element $\mathcal{F}_{M,mn}$ specifies the inner product between the directions along which the undriven steady state $\hat{\rho}_{\boldsymbol{0}}$ is displaced by the $m$th and $n$th drives. The QFIM can thus be used to introduce a measure of distance between the driven and undriven steady states, viz., the infinitesimal Bures distance  ~\cite{braunstein1994PRL,liu2020QuantumFisher}
\begin{eqnarray}    d_B^2(\hat{\rho}_{\boldsymbol{0}},\hat{\rho}_{\boldsymbol{\eps}}) = \frac{1}{4}\boldsymbol{\eps}^T \boldsymbol{\mathcal{F}}_M \boldsymbol{\eps} = \frac{1}{4}\sum_{j=1}^{M} \lambda_j \td{\eps}_j^2,
\label{eqn:bures}
\end{eqnarray}
where $\{\lambda_j\}$ are the eigenvalues of $\boldsymbol{\mathcal{F}}_M$, $\td{\boldsymbol{\eps}}=\boldsymbol{V}\boldsymbol{\eps}$ and $\boldsymbol{V}$ is the matrix of eigenvectors of $\boldsymbol{\mathcal{F}}_M$. The distance contribution of each eigendrive adds in quadrature and hence the induced responses define mutually orthogonal directions, or modes of synchronization, in parameter space. 

This notion of drive orthogonality is not merely an abstract information-geometric concept, rather, it has an operational meaning in QM. The precision limit for estimating an amplitude $\eps_m$ in the absence of all the other drives is  $(\Delta\eps_m)^2_\text{ab.}=1/(K\mathcal{F}_{M,mm})$. On the other hand, in their presence, their unknown (but small) amplitudes act as nuisance parameters~\cite{suzuki2020QuantumState} that generally degrade the ability to estimate $\eps_m$, with the corresponding precision limit given by $(\Delta\eps_m)^2_\text{pr.}=[\boldsymbol{\mathcal{F}}_M^{-1}]_{mm}/K$. For each drive $m$, we introduce an orthogonality measure 
\begin{eqnarray}
    D_m = \frac{(\Delta\eps_m)^2_\text{ab.}}{(\Delta\eps_m)^2_\text{pr.}} = \frac{1}{\mathcal{F}_{M,mm} [\boldsymbol{\mathcal{F}}_M^{-1}]_{mm}},
    \label{eqn:dmet}
\end{eqnarray}
such that $0\leq D_m \leq 1$. A value of $D_m=1$ implies that the response induced by drive $m$ is in a direction orthogonal to that of all the other drives in parameter space; this corresponds to the situation when $\mathcal{F}_{M,mn}=0 \;\forall \; n\neq m$ and hence the distance contribution of drive $m$ adds in quadrature in Eq.~(\ref{eqn:bures}). In particular, when all the independent drives are taken to be eigendrives, $D_m=1\;\forall\; m$. Conversely,  $D_m<1$ implies that the response induced by drive $m$ is partially indistinguishable from that of the other drives. A measure like Eq.~(\ref{eqn:dmet}) can guide the choice of drives to probe a many-body system. It can also reveal surprising features in the many-body synchronization response that emerge due to the interplay of gain, loss and interactions, as we illustrate below. 

\begin{figure}[tb]
    \centering
    \includegraphics[width=\columnwidth]{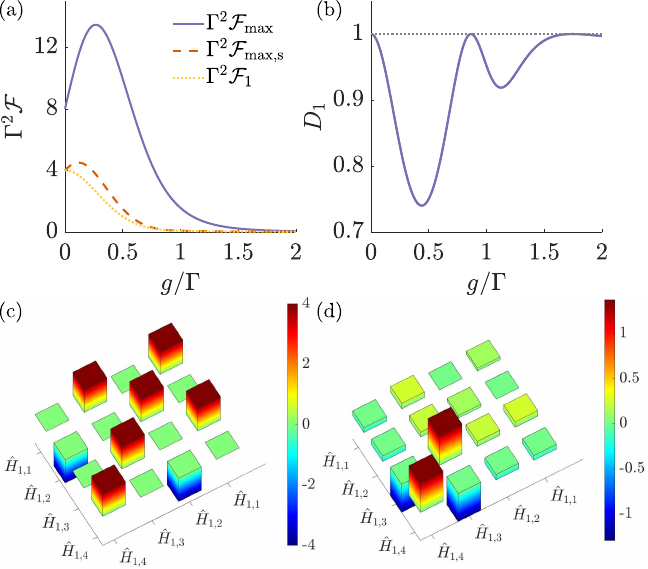}
    \caption{\textbf{Quantum synchronization under multiple drives in a two-qubit oscillator.} (a) Optimized QFI versus qubit-qubit coupling strength $g$, when the applied drive is optimized over the full set of $x$ drives ($\mathcal{F}_{\rm max}$) and when the set is restricted to unconditioned drives ($\mathcal{F}_{\rm max, s}$). The QFI when applying an unconditioned drive on only one of the qubits is also shown ($\mathcal{F}_1$). (b) Orthogonality measure [Eq.~(\ref{eqn:dmet})] for the unconditioned drives $\hat{H}_{1,1}$ and $\hat{H}_{1,2}$ [see Eq.~(\ref{eqn:drive_set})] versus $g$. (c) The QFIM $\boldsymbol{\mathcal{F}}_M$ for the full set of $x$ drives at $g=0$, and (d) at $g=\sqrt{3}\Gamma/2$. Other system parameters for these plots are described in the text.}
    \label{fig:tqo}
\end{figure}

\emph{Example: Two-qubit oscillator.}---As a minimal many-body oscillator, we consider a system of two coupled qubits subjected to local gain and damping channels. The gain (damping) of each qubit is described by operators  $\hat{O}_{g,j}=\sqrt{w_j}\hSig_j^+$ ($\hat{O}_{d,j}=\sqrt{\gamma_j}\hSig_j^-$). Here, $\hSig_j^\pm$ are the Pauli raising and lowering operators for qubit $j$.  We assume that the two qubits have identical frequencies $\omega_q$. In a frame rotating at $\omega_q$, the intrinsic system Hamiltonian is given by $\hat{H}_0 = -ig(\hSig_1^+\hSig_2^- - \hSig_2^+\hSig_1^-)$. The $U(1)$ symmetry operator for this system is $\hat{U}(\phi)=e^{-i\phi \hat{S}_z}$, where $\hat{S}_z=(\hSig_1^z+\hSig_2^z)/2$.

We consider synchronizing drives resonant with $\omega_q$. A total of $8$ linearly independent Hamiltonian drives can be applied, that break the $U(1)$ symmetry by coupling states that differ by $\Delta S_z = \pm 1$. These are 
\begin{eqnarray}
    \hat{\boldsymbol{H}}_1 = \{ \hSig_1^x, \hSig_2^x, \hSig_2^z \hSig_1^x, \hSig_1^z \hSig_2^x, \hSig_1^y, \hSig_2^y, \hSig_2^z \hSig_1^y, \hSig_1^z \hSig_2^y\}.
    \label{eqn:drive_set}
\end{eqnarray}
The drives $\hat{H}_{1,m},m=1,2,5,6$ are unconditioned drives while the remaining ones are conditioned on the state of the other qubit. All drives have the same magnitude as quantified by the Frobenius norm $\lVert \hat{O}\rVert_F^2 = {\rm{Tr}}[\hat{O}^\dag\hat{O}]$ and satisfy trace orthonormality, i.e., ${\rm Tr}[\hat{H}_{1,m}\hat{H}_{1,n}] \propto \delta_{m,n}$. This ensures that any linear combination $\hat{H}_{\boldsymbol{n}} = \boldsymbol{n}\cdot \hat{\boldsymbol{H}}_{1}$ has the same magnitude, and thus enhancements in QS obtained by drive optimization are not due to arbitrary scale factors. Due to the symmetry of the system, the $8\times 8$ QFIM is block diagonal, with each block given by identical $4\times 4$ matrices, respectively corresponding to drives along $x$ and $y$~\cite{supp}. Hence, we only analyze the $x$ drives $\hat{H}_{1,m}, m=1,\ldots,4$ below, denoting their $4\times4$ QFIM by $\boldsymbol{\mathcal{F}}_M$.

We consider the case where $\gamma_1>w_1$ and $\gamma_2<w_2$. The latter corresponds to an effective negative temperature bath for qubit $2$, which can be engineered, e.g., using an auxiliary level~\cite{vaidya2024exploring}. We choose $\gamma_1=w_2=\Gamma$ and $w_1=\gamma_2=0$, although many of our conclusions are not restricted to this particular parameter set. Figure~\ref{fig:tqo}(a) shows $\mathcal{F}_{\rm{max}}$, the maximum eigenvalue of $\boldsymbol{\mathcal{F}}_M$, as a function of the qubit-qubit coupling $g$. Interestingly, we observe that the peak value of $\mathcal{F}_{\rm{max}}$ occurs at non-zero $g$. Although the system consists of only two qubits, this feature points to the potential to enhance QS by engineering interactions in ensembles of self-sustained units, that could thus serve as highly sensitive many-body dissipative quantum sensors. For larger $g$, $\mathcal{F}_{\rm{max}}$ decreases monotonically as the energy levels shift appreciably, making the external drives off-resonant with any pair of levels.

In practice, it may be challenging to apply conditioned drives such as $\hat{H}_{1,3},\hat{H}_{1,4}$. A power of our approach is that we can now analyze QS under unconditioned drives alone by considering the $2\times 2$ submatrix $\boldsymbol{\mathcal{F}}_{M,\text{s}}$ of $\boldsymbol{\mathcal{F}}_M$ that corresponds to $\hat{H}_{1,1},\hat{H}_{1,2}$. The orange line in Fig.~\ref{fig:tqo}(a) shows $\mathcal{F}_{\rm{max, s}}$, the maximum eigenvalue of $\boldsymbol{\mathcal{F}}_{M,\text{s}}$, which is significantly lower than $\mathcal{F}_{\rm{max}}$ over the entire range of $g$. This highlights the role of the conditioned drives in boosting QS in this system, which is also reflected in the composition of the optimal drives, i.e., eigenvectors~\cite{supp}. Nevertheless, $\mathcal{F}_{\rm{max, s}}$ also peaks at non-zero $g$, implying that it is enhanced by qubit-qubit coupling. Finally, the yellow line shows $\mathcal{F}_1$, the QFI when driving only either one of the two qubits. Notably, $\mathcal{F}_1$ decreases monotonically and does not show any enhancement at non-zero $g$. These results show that our approach can unlock the full QS potential of many-body quantum oscillators by systematically identifying the optimal drives to apply to these systems. 

Restricting our analysis to $\boldsymbol{\mathcal{F}}_{M,\text{s}}$, we now study the mutual orthogonality of the responses induced by $\hat{H}_{1,1}$ and $\hat{H}_{1,2}$. In Fig.~\ref{fig:tqo}(b), we plot $D_1$ [Eq.~(\ref{eqn:dmet})] as $g$ is varied. Note that $D_1=D_2$ always for a $2\times2$ QFIM. At $g=0$, the two qubits are uncoupled and $D_1=1$ since the coherences established by $\hat{H}_{1,1}$ and $\hat{H}_{1,2}$ are localized on the respective qubits. For nonzero $g$, driving either qubit establishes coherences that are delocalized over the two-qubit system. Hence, the responses induced by $\hat{H}_{1,1}$ and $\hat{H}_{1,2}$ in general become partially indistinguishable, leading to $D_1<1$.

Interestingly, at the particular point $g\equiv g^*=\sqrt{3}\Gamma/2$, the system exhibits a revival of perfect orthogonality ($D_1=1$). This behavior is remarkable, because the two qubits are not decoupled and yet driving either qubit induces a response that is perfectly distinguishable from driving the other qubit. Unlike at $g=0$, the perfect orthogonality of the two drives at this point is thus a many-body effect. This is immediately evident in the QFIM for the full set of $x$ drives, $\boldsymbol{\mathcal{F}}_M$, which we plot at $g=0$ and $g=g^*$ in Figs.~\ref{fig:tqo}(c) and (d). In both cases, the perfect orthogonality of the unconditioned drives is evident in the vanishing off-diagonal element between $\hat{H}_{1,1}$ and $\hat{H}_{1,2}$. However, at $g=0$, there are no nonzero off-diagonal elements between \emph{any} two drives--- unconditioned or conditioned--- that act on different qubits. On the other hand, at $g=g^*$, we observe nonzero off-diagonal elements between, e.g., $\hat{H}_{1,1}$ and $\hat{H}_{1,4}$, and between $\hat{H}_{1,2}$ and $\hat{H}_{1,3}$. This implies that, at $g=g^*$, the coherences established by drives on different qubits are not in general localized on the respective qubits. An analysis of density matrix elements reveals that the orthogonality of $\hat{H}_{1,1}$ and $\hat{H}_{1,2}$ at $g=g^*$ results from a blockade in coherence buildup between specific eigenstates of $\hat{\rho}_0$~\cite{supp} when either drive is applied. Hence, the QFIM can serve as a  powerful tool to capture intriguing many-body effects in QS that may not be evident otherwise.

\emph{Conclusion and Outlook.}---We have shown that the QFI framework, an established toolbox of QM, has versatile applications in the study of QS. The connection between QS and dissipative quantum sensing demonstrated here has both fundamental and practical relevance: On the one hand, the QFI framework can be used, e.g., to guide the choice of drives to probe the system, and to unravel many-body QS effects. On the other, our work opens the domain of quantum sensing as a broad application area for QS.  An exciting research direction is the use of quantum synchronizing systems with engineered gain and loss to surpass precision limits imposed by unavoidable decoherence mechanisms~\cite{reiter2017NatComm, xie2020PRApp}. 
Furthermore, our work generalizes straightforwardly to the study of mutual synchronization of quantum systems, by replacing the synchronizing drive with a weak coupling between one or more quantum self-sustained oscillators~\cite{ameri2015PRA,roulet2018quantum}. Additionally, it will be interesting to extend this approach to explore QS under finite driving strengths~\cite{walter2014quantum}.

\begin{acknowledgments}
We thank Jarrod Reilly, Samarth Hawaldar, Arvind Mamgain and Baladitya Suri for discussions and comments on the manuscript. We acknowledge the use of QuTiP for numerical results~\cite{johansson2012qutip}. S.B.J. acknowledges support from the Deutsche Forschungsgemeinschaft (DFG, German Research Foundation) through projects A4 and A5 in TRR-185 “OSCAR”. A.S. acknowledges the support of a C.V. Raman Post-Doctoral Fellowship, IISc.
\end{acknowledgments}

\providecommand{\noopsort}[1]{}\providecommand{\singleletter}[1]{#1}%

\end{document}


\newcommand{\iitm}{\affiliation{Department of Physics, Indian Institute of Technology Madras, Chennai, India, 600036.}}
\newcommand{\iisc}{\affiliation{Department of Instrumentation and Applied Physics, Indian Institute of Science, Bangalore, India, 560012.}}
\newcommand{\optimas}{\affiliation{Physics Department and Research Center OPTIMAS, University of Kaiserslautern-Landau, D-67663, Kaiserslautern, Germany}}

\title{Supplemental Material: Quantum Synchronization and Dissipative Quantum Sensing}

\author{Gaurav M. Vaidya}
\iitm
\author{Simon B. J\"ager}
\optimas
\author{Athreya Shankar}
\iisc

\date{\today}

\maketitle

\tableofcontents

\section{QFI based QS measure for a generic perturbation $\lv_1$}

In the main text, we assumed that the perturbation $\lv_1$ only introduces coherences between $\ket{k},\ket{k'}$ belonging to different $U(1)$ symmetry sectors, which allowed us to use Eq.~(4) as our QS measure. However, a general perturbation can additionally introduce (i) changes in the population of each level $\ket{k}$ and (ii) coherences between $\ket{k},\ket{k'}$ in the same $U(1)$ symmetry sector. A QS measure must be sensitive only to the coherences breaking the $U(1)$ symmetry and should not be sensitive to the above two effects. Here, we discuss how the QFI, Eq.~(4), must be modified to construct a QS measure for a general perturbation $\lv_1$. 

To do so, we first formally define what we mean by $U(1)$ symmetry sectors. For a symmetry operator given by $\hat{U}(\phi)=e^{-i\phi\hat{O}}$, with $\hat{O}$ a Hermitian operator, we denote the eigenvalues by the functions $g_\mu(\phi),\mu=1,\ldots$. For each eigenvalue, there are in general a set $S_\mu$ of orthogonal eigenvectors, such that 
\begin{equation}
    S_\mu = \{\ket{\chi}\; :\; \hat{U}(\phi)\ket{\chi} = g_\mu(\phi)\ket{\chi} \}.
\end{equation}
Hence, the indices $\mu$ define eigenvectors belonging to different $U(1)$ symmetry sectors. We will use the notation $\ket{\chi}_{\mu\nu},\nu=1,2,\ldots$ to index eigenvectors in the set $S_{\mu}$.

We assume that the intrinsic dynamics of the system $\lv_0$ and the undriven steady state $\hat{\rho}_0$ are $U(1)$ symmetric. For the latter, this means that $\hat{U}(\phi)\hat{\rho}_0 \hat{U}^\dag(\phi) = \hat{\rho}_0$. Writing the spectral decomposition $\hat{\rho}_0 = \sum_k q_k \ket{k}\bra{k}$, the $U(1)$ symmetry implies that each eigenvector can be written as a linear combination of vectors in a single symmetry sector $S_\mu$:
\begin{equation}
\ket{k} = \sum_{\nu} c_{\nu}\ket{\chi}_{\mu\nu}.    
\end{equation}
Hence, each vector $\ket{k}$ is an eigenvector of $\hat{U}(\phi)$ with eigenvalue $g_\mu(\phi)$ for some fixed $\mu$.

For the superoperator $\lv_0$, the $U(1)$ symmetry implies that it commutes with the superoperator $\hat{U}(\phi)\cdot \hat{U}^\dag(\phi)$. In particular, this implies that 
\begin{equation}
    \hat{U}(\phi) (\lv_0 \ket{\chi}_{\mu\nu}\bra{\chi}_{\mu'\nu'}) \hat{U}^\dag(\phi) = \lv_0 (\hat{U}(\phi)\ket{\chi}_{\mu\nu}\bra{\chi}_{\mu'\nu'}\hat{U}^\dag(\phi)) = g_\mu(\phi)g_{\mu'}(\phi) (\lv_0 \ket{\chi}_{\mu\nu}\bra{\chi}_{\mu'\nu'}).
    \label{eqn:l0_u1}
\end{equation}
This implies that $\lv_0 \ket{\chi}_{\mu\nu}\bra{\chi}_{\mu'\nu'}$ is also a Liouville space eigenvector of $\hat{U}(\phi)\cdot \hat{U}^\dag(\phi)$, i.e., $\lv_0$ can only couple a Liouville basis state $\ket{\chi}_{\mu\nu}\bra{\chi}_{\mu'\nu'}$ with other Liouville basis states $\ket{\chi}_{\td{\mu}\td{\nu}}\bra{\chi'}_{\td{\mu}'\td{\nu}'}$ which share the same eigenvalue for the superoperator $\hat{U}(\phi)\cdot \hat{U}^\dag(\phi)$, i.e. $g_{\td{\mu}}(\phi)g_{\td{\mu}'}(\phi) = g_{\mu}(\phi)g_{\mu'}(\phi)\; \forall\;\phi$.

Given a general perturbation $\lv_1$, we can write it as $\lv_1 = \lv_{1,s} + \lv_{1,n}$, where $\lv_{1,s}$ is the $U(1)$ symmetric part, i.e. it only couples Liouville basis states with the same eigenvalue, similar to $\lv_0$ above, and $\lv_{1,n}$ is the part that is not $U(1)$ symmetric and \emph{only} couples Liouville basis states with different eigenvalues. Using linearity of superoperator action and the $U(1)$ symmetry properties of $\hat{\rho}_0$ and $\lv_0$, we can show that Eq.~(2) can be expressed as 
\begin{equation}
    \partial_{\eps}\hat{\rho} \equiv \partial_{\eps}\hat{\rho}_{s} + \partial_{\eps}\hat{\rho}_{n} = -\lv_0^{-1}\lv_{1,s}\hat{\rho}_0 -\lv_0^{-1}\lv_{1,n}\hat{\rho}_0, 
\end{equation}
where the differential operator $\partial_{\eps}\hat{\rho}_s$ is $U(1)$ symmetric and hence its matrix elements satisfy 
\begin{equation}
    \braket{k|\partial_{\eps}\hat{\rho}_s|k'} \neq 0\; \implies \ket{k},\ket{k'}\in \mathrm{span}(S_{\mu}),
\end{equation}
for some value of $\mu$, i.e. $\ket{k},\ket{k'}$ belong to the same $U(1)$ symmetry sector. On the other hand, the matrix elements of  $\partial_{\eps}\hat{\rho}_n$ satisfy 
\begin{equation}
    \braket{k|\partial_{\eps}\hat{\rho}_n|k'} \neq 0\; \implies \ket{k}\in \mathrm{span}(S_{\mu}) \;{\rm and}\; \ket{k'}\in \mathrm{span}(S_{\mu'})
\end{equation}
for some $\mu,\mu'$ such that $\mu \neq \mu'$, i.e. $\ket{k},\ket{k'}$ belong to different $U(1)$ symmetry sectors.

Using the above observations on the matrix elements, the QFI, Eq.~(4), can be expressed as 
\begin{equation}
    \mathcal{F} \equiv \mathcal{F}_s + \mathcal{F}_n = 
    \sum\limits_\mu     \sum\limits_{\substack{k,k'\\q_k+q_{k'}>0\\\ket{k},\ket{k'}\in \;\mathrm{span}(S_\mu)}} \frac{2}{q_k+q_{k'}}\abs{\braket{k'|\partial_\eps \hat{\rho}_s|k}}^2 +
    \sum\limits_{\substack{\mu,\mu'\\\mu\neq \mu'}}     \sum\limits_{\substack{k,k'\\q_k+q_{k'}>0\\\ket{k}\in \;\mathrm{span}(S_{\mu})\\\ket{k'}\in \;\mathrm{span}(S_{\mu'})}} \frac{2}{q_k+q_{k'}}\abs{\braket{k'|\partial_\eps \hat{\rho}_n|k}}^2.    
\end{equation}
Hence, we can write $\mathcal{F}$ as the sum of the QFI $\mathcal{F}_s$ resulting from the $U(1)$ symmetric perturbation $\lv_{1,s}$ and the QFI $\mathcal{F}_n$ resulting from the part $\lv_{1,n}$ that \emph{only} couples different $U(1)$ symmetry sectors. Hence, we can discard $\mathcal{F}_s$ and use $\mathcal{F}_n$ as the QS measure. Physically, it reflects the sensitivity to the amplitude $\eps$ if $\lv_{1,n}$ alone was applied to drive the system. 

In the presence of multiple drives, it is straightforward to extend the above arguments to show that the entire QFI matrix can be expressed as 
\begin{equation}
    \boldsymbol{\mathcal{F}}_M = \boldsymbol{\mathcal{F}}_{M,s} + \boldsymbol{\mathcal{F}}_{M,n}, 
\end{equation}
where $\boldsymbol{\mathcal{F}}_{M,s}$ and $\boldsymbol{\mathcal{F}}_{M,n}$ are evaluated from Eq.~(4) of the main text using the differential operators $\{\partial_{\eps_m}\hat{\rho}_s\}$ and $\{\partial_{\eps_m}\hat{\rho}_n\}$ respectively in place of the complete differential operators $\{\partial_{\eps_m}\hat{\rho}\}$. Analysis of the QFI matrix $\boldsymbol{\mathcal{F}}_{M,n}$ then amounts to analyzing the multiple drive QS scenario by replacing each drive with the corresponding component that only couples different $U(1)$ symmetry sectors.

\section{Quantum van der Pol oscillator: Classical limit}

\subsection{Convergence of $\td{\Omega}\to \mathcal{F}/2$}
\label{sec:convergence}
In this section we demonstrate that the synchronization measure $\tilde{\Omega}$ becomes equal to $\mathcal{F}/2$ in the classical limit. Our analysis is more general than the one used in the Letter. In general, we consider an oscillator whose intrinsic dynamics $\lv_0$ leads to a steady state $\hat{\rho}_0$ that is diagonal in the Fock basis $\ket{m}_0$ with $m=0,1,\dots$. The Fock basis plays here the role as the eigenbasis of the underlying $U(1)$ symmetry. Under a weak external drive, the new steady state is $\hat{\rho}$. In this situation, the entropy based QS measure is defined as~\cite{jaseem2020generalized} 
\begin{align}
    \Omega = S(\hat{\rho}_{\rm diag}) - S(\hat{\rho}), 
\end{align}
where $S(\hat{\rho})$ is the von Neumann entropy of a density matrix $\hat{\rho}$. The  density matrix $\hat{\rho}_{\rm diag}$ is obtained by deleting all off-diagonal elements of $\hat{\rho}$, when it is expressed in the eigenbasis of the undriven steady state $\hat{\rho}_0$. Note that $\hat{\rho}_\mathrm{diag}$ is not just the undriven steady state, since it contains the modified populations under the action of the external drive. In the following, we will derive an analytical expression for $\Omega$ using second order perturbation theory and show that it is proportional to the QFI when the steady state populations and coherences vary smoothly and slowly across the ladder of states, which is true in the classical limit of the quantum van der Pol oscillator.

We consider the effect of a resonant, weak external drive $\eps \hat{H}_1 = \eps \hat{x}$ on the oscillator. The idea is now to derive $\Omega$ in second order in $\eps$. For this we write the Liouvillian as $\lv = \lv_0+\eps \lv_1$ and the steady state up to second order in $\eps$:
\begin{align}
	\hat{\rho}=\hat{\rho}_0+\eps \hat{\rho}_1+\eps^2\hat{\rho}_2.\label{eq:rho}
\end{align}
Solving for the steady state in each order in $\eps$, we get the following conditions:
\begin{align}
    \mathcal{L}_0\hat{\rho}_0=&0,\\
    \mathcal{L}_0\hat{\rho}_1=&-\mathcal{L}_1\hat{\rho}_0,\label{eq:L1}\\
    \mathcal{L}_0\hat{\rho}_2=&-\mathcal{L}_1\hat{\rho}_1.\label{eq:L2}
\end{align}
Since the unperturbed steady state is diagonal in the Fock basis $\ket{m}_0$, we can write 
\begin{align}
\hat{\rho}_0=&\sum_{m=0}^{\infty}a_{m}\ket{m}_0\bra{m}_0,
\end{align}
with $a_{m}>0$.
Using now the explicit form of $\hat{H}_1$, Eqs.~\eqref{eq:L1} and \eqref{eq:L2}, and the $U(1)$ symmetry properties of $\lv_0^{-1}$ [see Eq.~(\ref{eqn:l0_u1})], we can write the forms of $\hat{\rho}_1,\hat{\rho}_2$ to be 
\begin{align}
	\hat{\rho}_1=&\sum_{m=0}^{\infty}(b_{m}\ket{m}_0\bra{m+1}_0 +b_{m}^*\ket{m+1}_0\bra{m}_0),\label{eq:rho1}\\
	\hat{\rho}_2=&\sum_{m=0}^{\infty}c_{m}\ket{m}_0\bra{m}_0+\sum_{m=0}^{\infty}d_{m}\ket{m}_0\bra{m+2}_0+\sum_{m=0}^{\infty}d_{m}^*\ket{m+2}_0\bra{m}_0,\label{eq:rho2}
\end{align}
where the coefficients $b_m,d_m$, are in general complex numbers and $c_m$ are real. With this, we have already found the diagonal representation
\begin{align}
    \hat{\rho}_{\mathrm{diag}}=\sum_{m=0}^{\infty}(a_m+\epsilon^2c_m)\ket{m}_0\bra{m}_0,\label{eq:diag}
\end{align}
which can be used to calculate $S(\hat{\rho}_{\mathrm{diag}})$. 

Next, we find the diagonal representation of $\hat{\rho}$. For this, we write the eigenvectors of $\hat{\rho}$ as a series in $\eps$  
\begin{align}
\ket{m} =\ket{m}_0 + \eps\ket{m}_1 + \eps^2\ket{m}_2,\label{eq:m}
\end{align}
and accordingly write the eigenvalues as 
\begin{align}
\lambda_m=\lambda_{m,0}+\eps\lambda_{m,1}+\eps^2\lambda_{m,2}.\label{eq:lambda}
\end{align}
Using the fact that the vectors~\eqref{eq:m} are the eigenvectors of Eq.~\eqref{eq:rho} with eigenvalues~\eqref{eq:lambda}, we can derive the following conditions using perturbation theory
\begin{align}
	\hat{\rho}_0 \ket{m}_0=&\lambda_{m,0} \ket{m}_0, \nonumber\\	\hat{\rho}_0 \ket{m}_1+\hat{\rho}_1 \ket{m}_0=&\lambda_{m,0}\ket{m}_1 + \lambda_{m,1}\ket{m}_0, \nonumber\\ \hat{\rho}_0\ket{m}_2+\hat{\rho}_1\ket{m}_1+\hat{\rho}_2\ket{m}_0=&\lambda_{m,0}\ket{m}_2+\lambda_{m,1}\ket{m}_1+\lambda_{m,2}\ket{m}_0.
	\end{align}

Note that $\hat{\rho}_1$ in Eq.~\eqref{eq:rho1} is off-diagonal in $\ket{m}_0$. Hence, we have
\begin{align}
	\lambda_{m,1}=&0,\\
	\ket{m}_1=&[\lambda_{m,0}-\hat{\rho}_0]^{-1}\hat{\rho}_1\ket{m}_0.
\end{align}
Furthermore, using Eq.~\eqref{eq:rho2} we can calculate
\begin{align}
	\lambda_{m,2}=&\braket{m|_0\hat{\rho}_1|m}_1 +\braket{m|_0\hat{\rho}_2|m}_0
	=\braket{m|_0\hat{\rho}_1[\lambda_{m,0}-\hat{\rho}_0]^{-1}\hat{\rho}_1|m}_0+\braket{m|_0\hat{\rho}_2|m}_0.
\end{align}
This results in
\begin{align}
    \lambda_{0,2}=&\frac{|b_0|^2}{a_0-a_1}+c_0,\\
    \lambda_{m,2}=&\frac{|b_m|^2}{a_m-a_{m+1}}+\frac{|b_{m-1}|^2}{a_m-a_{m-1}}+c_m,\text{ for } m>0.
\end{align}
Summarizing the results, we find the following expressions for the eigenvalues of $\hat{\rho}$: 
\begin{align}
	\lambda_0=&a_0+\eps^2\frac{|b_0|^2}{a_0-a_{1}}+\eps^2 c_0,\nonumber\\
	\lambda_m=&a_m+\eps^2\frac{|b_m|^2}{a_m-a_{m+1}}+\eps^2\frac{|b_{m-1}|^2}{a_m-a_{m-1}}+\eps^2 c_m,\text{ for }m>0.
\end{align}

We are now ready to calculate $\Omega$. We get 
\begin{align}
	\Omega = S(\rho_{\mathrm{diag}})-S(\rho)=&-\sum_m\left[[a_m+\eps^2 c_m]\log\left[a_m+\eps^2 c_m\right]-\lambda_m\log(\lambda_m)\right]\nonumber\\
	\approx&\sum_m[\lambda_m-a_m-\eps^2 c_m]\left[\log\left(a_m\right)+1\right]\nonumber\\
		=&\eps^2 \sum_{m=0}^{\infty}|b_m|^2\frac{\log(a_m)-\log(a_{m+1})}{a_m-a_{m+1}},
	\end{align}
where in the approximation step, we have used $(x_0+x_1)\log(x_0+x_1)\approx x_0\log(x_0)+x_1[\log(x_0)+1]$ for $x_1\ll x_0$, with $x_1=\lambda_m-(a_m+\eps^2 c_m)$ and $x_0 = a_m+\eps^2 c_m$.

If $a$ and $b$ are smooth and slowly varying functions of $m$, then we can replace the finite difference in the summand with the derivative $d \log(a_m)/d a_m = 1/a_m$ to obtain 
\begin{align}
	\td{\Omega} = \frac{\Omega}{\eps^2} \approx &\sum_{m=0}^{\infty}\frac{|b_m|^2}{a_m}.
 \label{eqn:omega_classical}
\end{align}

On the other hand, for the QFI, we need to evaluate 
\begin{align}
\mathcal{F}=\sum\limits_{\substack{m,m'\\a_m+a_{m'}>0}} \frac{2}{a_m+a_{m'}}\abs{\braket{m'|_0\hat{\rho}_1|m}_0}^2=\sum_{m=0}^{\infty}\frac{4|b_{m}|^2}{a_m+a_{m+1}} \approx \sum_{m=0}^{\infty}\frac{2|b_{m}|^2}{a_m},
\label{eqn:qfi_classical}
\end{align}
where the last approximation is justified if $a_m,b_m$ are smooth and slowly varying with $m$. Hence, we see that $\td{\Omega}\to \mathcal{F}/2$ under fairly general assumptions of smooth variation of populations and coherences across the ladder of Fock states. In particular, these assumptions are valid in the classical limit of the quantum vdP oscillator, where we observe in Fig.~1 of the main text that $\td{\Omega}\to \mathcal{F}/2$ for $\kappa_1/\kappa_2\gg 1$.

\subsection{Evaluation of $\mathcal{F}$ and $\mu$ in classical limit}

To evaluate $\mathcal{F}$ and $\mu$ in the classical limit of the quantum vdP oscillator, we use the results presented in Sec.~SII D of the Supplemental Material of Ref.~\cite{dutta2019PRL}. We briefly summarize the relevant results using our notation. 

First, in the limit that $\kappa_1\gg\kappa_2$, the occupations $\{a_m\}$ of the Fock states $\{\ket{m}_0\}$ are sharply peaked about the classical expectation value $\mbar = \kappa_1/(2\kappa_2)\gg 1$. This distribution can be approximated as a continuous function $u(y)$ of the variable $y=(m-\mbar)/\sqrt{2\mbar}$, which is found to be 
\begin{align}
    u(y) = \frac{1}{\sqrt{2\mbar}}\sqrt{\frac{2}{3\pi}} e^{-2y^2/3}. 
    \label{eqn:uy}
\end{align}
This distribution is normalized according to the condition $\sum_m a_m=1$, which amounts to  $\sqrt{2\mbar}\int dy u(y) \approx 1$. Equation~(\ref{eqn:uy}) implies that the steady state number distribution has a mean $\mbar$ and standard deviation given by $\sqrt{3\mbar/2}$.

Second, in the presence of a weak drive, the coherences between adjacent levels, $b_m$, can also be approximated by a continuous distribution $v(y)$. An important observation of Ref.~\cite{dutta2019PRL} is that this distribution has the same functional form as $u(y)$ and is given by 
\begin{equation}
    v(y) = -\frac{2i\eps}{\kappa_1}\frac{1}{\sqrt{27\pi}}e^{-2y^2/3}.
    \label{eqn:vy}
\end{equation}
We note that the above equation differs by a factor of $2i$ from Eq.~(S37) of Ref.~\cite{dutta2019PRL} because of the different definitions of the linear drive. We also remind that because of the different notation, the symbol $\eps$ is used to represent different quantities in this work and Ref.~\cite{dutta2019PRL}.

Using the above two results for $u(y), v(y)$, we can use the continuum approximation of Eq.~(\ref{eqn:qfi_classical}) to evaluate $\mathcal{F}$. This leads to 
\begin{align}
    \mathcal{F} \approx  \sum_{m=0}^{\infty}\frac{2|b_{m}|^2}{a_m} \approx 2 \sqrt{2\mbar}\int_{-\sqrt{\mbar/2}}^\infty  \frac{\abs{v(y)}^2}{u(y)} \approx 2 \sqrt{2\mbar}\int_{-\infty}^\infty  \frac{\abs{v(y)}^2}{u(y)} = \frac{4}{9\kappa_1\kappa_2},  
\end{align}
where in the second-to-last step, we have extended the lower limit of integration to $-\infty$ since $\mbar\gg 1$, resulting in a standard gaussian integral.

In order to find $\mu=(\partial_\eps \ev{\hat{p}})^2/{\rm Var}(\hat{p})$, we separately evaluate the numerator and denominator. The denominator is straightforwardly given as ${\rm Var}(\hat{p}) = 1/4+\mbar/2\approx \mbar/2$. To find the numerator, we use Eq.~(\ref{eqn:vy}) to evaluate 
\begin{align}
    \ev{\hat{a}} = \sum_m \sqrt{m}b_m = \sqrt{2\mbar}\int_{-\sqrt{\mbar}/2}^{\infty} dy \sqrt{\sqrt{2\mbar}y + \mbar} \;v(y).
\end{align}
Since $v(y)\propto e^{-2y^2/3}$, it is significant only for $y\sim O(1)$, whereas $\mbar\gg 1$. Hence, we can approximate $\sqrt{\sqrt{2\mbar}y + \mbar}\approx \sqrt{\mbar}$. Once again extending the lower limit of the integral to $-\infty$ and performing the Gaussian integral, we obtain 
\begin{align}
    \ev{\hat{a}} \approx \frac{2\eps\mbar}{3\kappa_1}.
\end{align}
We can now evaluate the numerator of $\mu$, $(\partial_\eps\ev{\hat{p}})^2$, and arrive at the final result that 
\begin{align}
    \mu = \frac{(\partial_\eps\ev{\hat{p}})^2}{{\rm Var}(\hat{p})} = \frac{4\mbar^2}{9\kappa_1^2}\times \frac{2}{\mbar} = \frac{4}{9\kappa_1\kappa_2}.
\end{align}

\section{Quantum van der Pol oscillator: Quantum limit}

Here, we compute analytic expressions for $\mathcal{F}$ and $\mu$ in the regime $\kappa_1/\kappa_2\ll 1$. In the limit $\kappa_1/\kappa_2\to 0$, the analysis can be restricted to the lowest three levels, which respectively have populations~\cite{lee2013quantum} 
\begin{align}
    p_0=\frac{2}{3}, \; p_1=\frac{1}{3}, \; p_2=0.
\end{align}
Under the action of the drive $\eps\hat{x}$, the equations of motion for the coherences between these levels can be written as 
\begin{align}
    \frac{d}{dt}\rho_{10} &= -\frac{3}{2}\kappa_1 \rho_{10} -i\frac{\eps}{2}(p_1-p_0) + O\left(\frac{\kappa_1}{\kappa_2}\right), \nonumber\\
    \frac{d}{dt}\rho_{21} &= - \kappa_2 \rho_{21} + \sqrt{2}\kappa_1\rho_{10} -i\frac{\eps}{\sqrt{2}}(p_2-p_1) + O\left(\frac{\kappa_1}{\kappa_2}\right).
\end{align}
The above equations imply that the steady state coherences satisfy 
\begin{align}
    \rho_{10} = \frac{i\eps}{9\kappa_1}, \; \frac{\rho_{21}}{\rho_{10}} \propto \frac{\kappa_1}{\kappa_2} \ll 1.
\end{align}
Hence, we only need to consider the $\rho_{10}$ matrix element in the computation of $\mathcal{F}$ and the numerator of $\mu$. 

The resulting value of $\mathcal{F}$ is given by 
\begin{align}
    \mathcal{F} = 2 \times \frac{2}{p_0+p_1}\abs{\partial_\eps \rho_{10}}^2 = \frac{4}{81\kappa_1^2},
\end{align}
where the extra factor of $2$ arises because of the sum over $k=0,k'=1$ and $k=1,k'=0$ in Eq.~(3) of the main text.

The numerator of $\mu$ is given by
\begin{align}
    (\partial_{\eps}\ev{\hat{p}})^2 = \abs{\partial_\eps \rho_{10}}^2 = \frac{1}{81\kappa_1^2}.
\end{align}
The denominator is given by the variance of $\hat{p}$ in the undriven steady state, which can be evaluated as 
\begin{align}
    {\rm Var}(\hat{p}) = \ev{\hat{p}^2}-\ev{\hat{p}}^2 = \ev{\hat{p}^2} = \frac{1}{4}(2\ev{\hat{a}^\dag\hat{a}}+1) = \frac{5}{12}. 
\end{align}
Hence, we find the value of $\mu$ to be
\begin{align}
    \mu = \frac{1}{81\kappa_1^2}\times \frac{12}{5} = \frac{4}{135\kappa_1^2}.
\end{align}

Thus, $\mu<\mathcal{F}$ in the quantum limit $\kappa_1/\kappa_2\ll 1$. This means that, estimating the drive amplitude $\eps$ using the method of moments approach~\cite{pezze2018RMP} on the `signal' $\ev{\hat{p}}$ is not the optimal estimation strategy in this limit. We can understand why this approach is sub-optimal by analyzing the contribution of the lowest three quantum levels to the quantity ${\rm Var}(\hat{p})$. To do so, we write 
\begin{align}
    {\rm Var}(\hat{p}) = \ev{\hat{p}^2} = -\frac{1}{4} \left\langle\left(\ket{0}\bra{1} + \sqrt{2}\ket{1}\bra{2} - \ket{1}\bra{0} - \sqrt{2}\ket{2}\bra{1}\right)^2\right\rangle
    = \frac{1}{4}\left(p_0 + p_1 + 2 p_1 \right)=\frac{5}{12}.
\end{align}
The last term $2p_1$ in the parenthesis arises because of the contribution of the $\ket{2}\bra{1}$ and $\ket{1}\bra{2}$ operators to the overall variance of $\hat{p}$. 

Equipped with this insight, we now define a new operator $\hSig_y$, which is just the truncation of $\hat{p}$ to the lowest two levels. Defining $\td{\mu} = (\partial_{\eps}\ev{\hSig_y})^2/{\rm Var}(\hSig_y)$
we observe that the `signal', i.e. the numerator, remains the same as before, but the denominator no longer contains the extra noise contribution from $\ket{2}\bra{0}$. Hence, applying the method-of-moments estimation technique using the mean and variance of $\hSig_y$ instead of $\hat{p}$, the signal-to-noise ratio saturates the QFI, i.e. $\td{\mu}=\mathcal{F}=4/(81\kappa_1^2)$.

\section{Two-qubit oscillator: Symmetry properties of QFIM}
\label{sec:tqo_sym}

In the main text, we identified $8$ possible drives that can be applied to the two-qubit oscillator (TQO), that in principle lead to an $8\times 8$ QFIM. Here, we will show that, thanks to the symmetry of the system, this $8\times 8$ QFIM is block-diagonal in drives along $x$ and $y$, with each set of drives corresponding to identical $4\times 4$ QFIMs.

The master equation for the intrinsic dynamics of the TQO is given by 
\begin{align}
    \frac{\partial\hat{\rho}}{\partial t} = \lv_0\hat{\rho} &= -i [\hat{H}_0,\hat{\rho}] + \sum_{j\in\{1,2\}} \gamma_j \mathcal{D}[\hSig_j^-]\hat{\rho} + \sum_{j\in\{1,2\}} w_j \mathcal{D}[\hSig_j^+]\hat{\rho}, \nonumber\\
    \hat{H}_0 &= -i g(\hSig_1^+\hSig_2^- - \hSig_2^+\hSig_1^-).
    \label{eqn:tqo_me_l0}
\end{align}
The $U(1)$ symmetry of the intrinsic dynamics means that $\lv_0$ and $\hat{\rho}_0$ are invariant under a unitary transformation by the operator $\hat{U}(\phi)=e^{-i\phi(\hSig_1^z+\hSig_2^z)/2}$. As a consequence, the steady state $\hat{\rho}_0$ has the form 
\begin{align}
    \hat{\rho}_0 = \sum_{q =1}^4 \lambda_q \ket{q}\bra{q},
\end{align}
where $\lambda_q$ are the (real, nonnegative) eigenvalues and the $\{\ket{q}\}$ are given by 
\begin{align}
    &\ket{1} = \ket{\downarrow\downarrow}, \nonumber\\
    &\ket{2} = a\ket{\downarrow\uparrow} + b\ket{\uparrow\downarrow}, \nonumber\\
    &\ket{3} = b\ket{\downarrow\uparrow} - a\ket{\uparrow\downarrow}, \nonumber\\
    &\ket{4} = \ket{\uparrow\uparrow}
    \label{eqn:rho0_eigvec}
\end{align}
where $a,b$ are real coefficients (see Sec.~\ref{sec:perfect_dist} for a discussion of why they are real).  The eigenvectors of $\hat{\rho}_0$ are also eigenvectors of $\hat{S}_z = (\hSig_1^z+\hSig_2^z)/2$, and hence transform under $\hat{U}(\phi)$ as 
\begin{align}
    \ket{1}\to e^{i\phi}\ket{1},\; \ket{2}\to\ket{2},\;\ket{3}\to\ket{3},\;\ket{4}\to e^{-i\phi}\ket{4}.
\end{align}

We now consider the action of a perturbative drive composed of a linear combination of the $4$ possible $x$ drives that changes $\hat{H}_0\to \hat{H}_0 + \boldsymbol{\eps}\cdot \hat{\boldsymbol{H}}_{x}$, where
\begin{align}
   \boldsymbol{\eps}\cdot \hat{\boldsymbol{H}}_{x} =&\eps_1 \hat{H}_{x,1} + \eps_2 \hat{H}_{x,2} + \eps_3 \hat{H}_{x,3} + \eps_4 \hat{H}_{x,4}:= \eps_1 \hSig_1^x + \eps_2 \hSig_2^x + \eps_3 \hSig_2^z\hSig_1^x + \eps_4 \hSig_1^z\hSig_2^x.
\end{align}
The corresponding change in the steady state is $\hat{\rho}_0\to \hat{\rho}_0 +\boldsymbol{\eps}\cdot \boldsymbol{\nabla}_{\boldsymbol{\eps}}\hat{\rho}$, where the vector of differential changes in $\hat{\rho}$ has elements given by 
\begin{align}
    \partial_{\eps_m} \hat{\rho} = i\lv_0^{-1}[\hat{H}_{x,m},\hat{\rho}_0].
\end{align}
Since the drives have non-zero matrix elements only between states differing by $\Delta S_z = \pm 1$, the perturbations to the steady state have the form 
\begin{align}
    \partial_{\eps_m} \hat{\rho} = c_{21}\ket{2}\bra{1} + c_{31}\ket{3}\bra{1} + c_{42}\ket{4}\bra{2} + c_{43}\ket{4}\bra{3} + \mathrm{H.c}.
\end{align}

The drives along $x$ can be transformed to a drive along any direction $\phi$ in the $x-y$ plane by the transformation $\boldsymbol{\hat{H}}_{\phi} = \hat{U}(\phi) \boldsymbol{\hat{H}}_{x}\hat{U}(\phi)^\dagger$. Here, $\phi=0$ ($\phi=\pi/2$) corresponds to drives along $x$ ($y$). Because of the $U(1)$ symmetry of $\lv_0$ and $\rho_0$, the $ \partial_{\eps_m} \hat{\rho}$ transform as 
\begin{align}
     (\partial_{\eps_m} \hat{\rho})_\phi =  i\lv_0^{-1}[\hat{U}_\phi\hat{H}_{x,m}\hat{U}_\phi^\dag,\hat{\rho}_0] = i \hat{U}_\phi \lv_0^{-1}[\hat{H}_{x,m},\hat{\rho}_0] \hat{U}_\phi^\dag = \hat{U}_\phi \partial_{\eps_m} \hat{\rho} \hat{U}_\phi^\dag.
\end{align}
Hence, the matrix elements of $\partial_{\eps_m} \hat{\rho}$ transform as 
\begin{align}
    \braket{k|(\partial_{\eps_m} \hat{\rho})_\phi|1} = e^{i\phi}\braket{k|\partial_{\eps_m} \hat{\rho}|1}, \;\braket{4|(\partial_{\eps_m} \hat{\rho})_\phi|k} = e^{i\phi}\braket{4|\partial_{\eps_m} \hat{\rho}|k},\; k=2,3.
    \label{eqn:drho_el_uphi}
\end{align}
As a result, the QFIM for the set of drives specified by $\hat{\boldsymbol{H}}_\phi$ is independent of $\phi$ as the phase factor cancels in the evaluation of the QFIM. This shows that the set of $x$ drives and the set of $y$ drives contribute identical $4\times 4$ QFI matrices to the full $8\times8$ QFIM. 

It now remains to show that the off-diagonal blocks, corresponding to driving one qubit along $x$ and the other along $y$, vanish. First, it can be shown by explicit calculation that the elements of the matrices $(\partial_{\eps_m} \hat{\rho})_{\phi=0} \; \forall \; m$, i.e. for all four drives along $x$, are purely imaginary (see Sec.~\ref{sec:perfect_dist} below Eq.~\eqref{eqn:tqo_ysol}). Consequently, from Eq.~\eqref{eqn:drho_el_uphi}, the matrix elements of all the  matrices $(\partial_{\eps_m} \hat{\rho})_{\phi=\pi/2}$, i.e., for driving along $y$, are purely real. Therefore, when evaluating the off-diagonal QFI matrix elements according to Eq.~(4) of the main text, the real part of the product of matrix elements of $(\partial_{\eps_m} \hat{\rho})_{\phi=0}$ and $(\partial_{\eps_m} \hat{\rho})_{\phi=\pi/2}$ vanishes. Hence, the QFIM is block diagonal in the $x$ and $y$ drives.

\section{Two-qubit oscillator: Optimized drives}
\label{sec:optimized_drives}
In Fig.~2 of the main text, we optimized the synchronization of the two-qubit oscillator and discussed the maximum QFI possible in the presence of the full set of $x$ drives or when the drives are limited to only the single-particle drives. Here, we discuss the optimal linear combinations of drives, i.e. the eigenvectors of the QFIM, that maximize the QFI.

\begin{figure}[!tb]
    \centering
    \includegraphics[width=\textwidth]{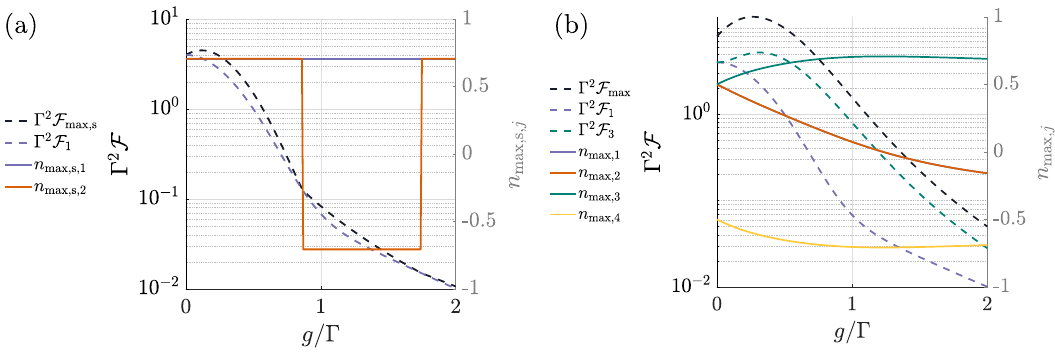}
    \caption{Optimized eigenvectors for synchronization of the two-qubit oscillator. (a) Drives optimized over the set of single-particle $x$ drives, and (b) Drives optimized over the full set of $x$ drives. In both cases, QFI values for various cases (see text) are shown as dashed lines, while the components of the optimized eigenvector that achieves $\mathcal{F}_{\rm max, s}$ or $\mathcal{F}_{\rm max}$} are shown as solid lines. The various curves and symbols are discussed in the text in Section~\ref{sec:optimized_drives}.
    \label{fig:tqo_eigvec}
\end{figure}

Figure~\ref{fig:tqo_eigvec}(a) plots $\mathcal{F}_{\rm max, s}$, the maximum QFI when the drives are restricted to be linear combinations of the single-particle drives $\hat{H}_{1,1}=\hSig_1^x$ and $\hat{H}_{1,2}=\hSig_2^x$ [see Eq.~(7) of the main text], as $g$ is scanned. For comparison, the QFI $\mathcal{F}_1$ when only either one of the two drives is applied is also shown. The components of the optimized eigenvector are shown on the same plot as solid lines. Because of the symmetry of the chosen gain and loss parameters for the two qubits, the $2\times2$ QFIM $\boldsymbol{\mathcal{F}}_{M,\mathrm{s}}$ has equal diagonal elements. Hence, the maximal eigenvector $\boldsymbol{n}_{\rm max, s}$ takes the form $(1,\pm 1)/\sqrt{2}$, where the sign depends on the sign of the off-diagonal element $\mathcal{F}_{M,\mathrm{s},12}$. This feature is evident in the solid lines, with the component $n_{{\rm max, s},1}$ corresponding to the amplitude of $\hSig_1^x$ fixed at $1/\sqrt{2}$ and $n_{{\rm max, s},2}$ displaying abrupt jumps between $\pm1/\sqrt{2}$. Physically, this implies that in certain ranges of the qubit-qubit coupling strength $g$, the qubits must be driven in phase to maximize the QFI whereas in other ranges they must be driven $180^\circ$ out-of-phase. The points where $n_{{\rm max}, s,2}$ changes sign corresponds to the values of $g$ where the off-diagonal element of $\boldsymbol{\mathcal{F}}_{M,\mathrm{s}}$ vanishes, making it proportional to the $2\times 2$ identity matrix. These are precisely the points where $D_1=D_2=1$ in Fig.~2(b) of the main text. Hence, zero-crossings of the off-diagonal element signify a crossover in the  optimal driving from in-phase to out-of-phase for the two qubits.

Figure~\ref{fig:tqo_eigvec}(b) extends the eigenvector analysis to the full set of $x$ drives, with $\hat{H}_{1,1}=\hSig_1^x,\hat{H}_{1,2}=\hSig_2^x,\hat{H}_{1,3}=\hSig_2^z\hSig_1^x$ and $\hat{H}_{1,4}=\hSig_1^z\hSig_2^x$. The black dashed line shows the variation of the maximal QFI, $\mathcal{F}_{\rm max}$ with $g$. Also shown for comparison are the QFI curves obtained under driving with only $\hat{H}_{1,1}$ ($\mathcal{F}_1$) and only $\hat{H}_{1,3}$ ($\mathcal{F}_3$). Because of the symmetric choice of bath parameters, $\mathcal{F}_2=\mathcal{F}_1$ and $\mathcal{F}_4=\mathcal{F}_3$.  Unlike in the case of single-particle drives, optimizing using the full set of $x$ drives leads to a smooth variation of the components of the maximal eigenvector, which are shown as solid lines. The component $n_{{\rm max},1}$ is identical to $n_{{\rm max},2}$ and is not visible since it is perfectly overlaid by the latter curve. For any value of $g$, we find that the optimal driving strategy is to apply the two single-particle drives in phase and with equal amplitude, and the two conditioned drives $\hat{H}_{1,3},\hat{H}_{1,4}$ out-of-phase but with equal amplitude. Furthermore, it is advantageous to assign larger amplitudes to the conditioned drives than to the single-particle drives for all values of $g$. Interestingly, there appears to be a special value of $g$ where the amplitudes of the single-particle drives vanish and maximum synchronization is obtained by driving the system with purely conditioned drives. The analysis of optimal eigenvectors thus promises to provide a window into the rich physics of the response of quantum many-body oscillators to external driving, and represents an exciting avenue for future work.  

\section{Two-qubit oscillator: Perfect drive orthogonality at non-zero $g$}
\label{sec:perfect_dist}

In this section, we will show by explicit calculation that, for the parameters considered in Fig.~2 of the main text, the choice of qubit-qubit coupling $g=\sqrt{3}\Gamma/2$ corresponds to a special point where the single-particle drives $\eps\hSig_1^x$ and $\eps\hSig_2^x$ lead to perfectly orthogonal responses, i.e. $D_1=D_2=1$ [see Eq.~(6) of the main text]. 

We first compute the steady-state solution of the undriven TQO for general choice of gain, loss and coupling rates. The only constraint we assume is that $w_j+\gamma_j=\Gamma$ for $j=1,2$, since this is sufficient for our analysis and simplifies the analytic expressions. Because of the $U(1)$ symmetry, the steady-state density matrix $\hat{\rho}_0$ is completely determined by the expectation values of four operators that are invariant under the $U(1)$ symmetry:
\begin{align}
    \boldsymbol{x}^T=(x_1,x_2,x_3,x_4) = \ev{\hat{\boldsymbol{x}}}^T = 
    (\ev{\hSig_1^z}, \ev{\hSig_2^z}, \ev{\hSig_1^+\hSig_2^-}, \ev{\hSig_1^z \hSig_2^z}).
\end{align}
Note that $\langle\hat{\sigma}_2^+\hat{\sigma}_1^-\rangle=x_3^*$ and expectation values of operators that are not invariant under the $U(1)$ symmetry are equal to zero:  $\langle\hat{\sigma}_j^{+}\rangle=0$ and $\langle\hat{\sigma}_{k}^z\hat{\sigma}_j^+\rangle=0$  for $j,k=1,2$ and $k\neq j$.

Writing down the equations of motion for these expectation values $\partial_t \boldsymbol{x} = {\rm Tr}[\hat{\boldsymbol{x}}\lv_0\hat{\rho}]$ and solving for the steady state solution, we find 
\begin{align}
    x_1 &= \frac{d_1 +2\gbar^2(d_1+d_2)}{1+4\gbar^2},\nonumber\\
    x_2 &= \frac{d_2 +2\gbar^2(d_1+d_2)}{1+4\gbar^2},\nonumber\\
    x_3 &= \frac{\gbar(d_1-d_2)}{2(1+4\gbar^2)},\nonumber\\
    x_4 &= \frac{d_1 d_2 + \gbar^2 (d_1+d_2)^2}{1+4\gbar^2},
    \label{eqn:tqo_xsol}
\end{align}
where $d_j = (w_j-\gamma_j)/\Gamma$ for $j=1,2$ and $\gbar=g/\Gamma$ are dimensionless parameters of the TQO. This calculation shows that, although $x_3$ can in general be complex, here it is purely real. The $4\times 4$ density matrix $\hat{\rho}_0$ can thus be entirely expressed using the above $4$ expectation values as a real, symmetric matrix. Hence, it has real eigenvalues and eigenvectors, which justifies the choice of real coefficients $a,b$ in Eq.~(\ref{eqn:rho0_eigvec}).

Under the influence of the single-particle drive $\eps\hSig_1^x$, operators that break $U(1)$ symmetry acquire non-zero expectation values. At $O(\eps)$, the operators acquiring non-zero steady-state expectation values are 
\begin{align}
    \boldsymbol{y}^T = \ev{\hat{\boldsymbol{y}}}^T = 
    ( \ev{\hSig_1^+}, \ev{\hSig_2^+}, \ev{\hSig_2^z\hSig_1^+}, \ev{\hSig_1^z\hSig_2^+} ),
\end{align}
which satisfy a set of linear equations $\boldsymbol{M} \boldsymbol{y} = i\eps\boldsymbol{c}$, where  
\begin{align}
   \boldsymbol{M}= 
    \begin{pmatrix}
    -\Gamma/2   &   0   &   0   &   g\\
    0   &   -\Gamma/2   &   -g  &   0\\
    \Gamma d_2  &   g   &   -3\Gamma/2  &   0\\
    -g  &   \Gamma d_1  &   0   &   -3\Gamma/2
    \end{pmatrix}, \;
    \boldsymbol{c} = 
    \begin{pmatrix}
    x_1\\
    0\\
    x_4\\
    -2 x_3
    \end{pmatrix}.
    \label{eqn:tqo_ysol}
\end{align}
Since $\boldsymbol{y}=i\eps \boldsymbol{M}^{-1}\boldsymbol{c}$, and $\eps, \boldsymbol{M}, \boldsymbol{c}$ are real, the elements of $\boldsymbol{y}$ are purely imaginary. It can be checked by explicit calculation of the source term $\boldsymbol{c}$, that the same argument applies when the drive is replaced by any arbitrary linear combination of the $4$ possible $x$ drives. Combined with the fact that the eigenvectors of $\hat{\rho}_0$ are all real, this proves the statement towards the end of Sec.~\ref{sec:tqo_sym} that the matrix elements of $(\partial_{\eps_m}\hat{\rho})_{\phi=0}$ are all purely imaginary. 

In Fig.~2 of the main text, we choose the system parameters $\gamma_1=w_2=\Gamma$ and $w_1=\gamma_2=0$. Equations~(\ref{eqn:tqo_xsol}) and~(\ref{eqn:tqo_ysol}) can then be used to find explicit expressions for $\boldsymbol{y}$ as a function of a single parameter $\gbar=g/\Gamma$. To find the specific point $\gbar^*$ where $D_1$ exhibits a revival to $1$, we are guided by  the structure of the numerically determined form of $\partial_\eps \hat{\rho}$ for the drive $\eps\hSig_1^x$. In particular, we observe a blockade in the buildup of certain coherences as we approach the revival point: The coherences established by the drive $\eps\hSig_1^x$ between the eigenstates $\ket{1}-\ket{3}$ and $\ket{1}-\ket{2}$ vanish [see Eq.~(\ref{eqn:rho0_eigvec})]. On the other hand, if the drive was replaced by $\eps\hSig_2^x$, the $\ket{4}-\ket{3}$ and $\ket{4}-\ket{2}$ coherences vanish. Hence, at the point $\gbar^*$ where $D_1=1$, the drives lead to perfectly orthogonal responses since they only induce coherences between mutually exclusive pairs of levels. 

To determine the point $\gbar^*$, we therefore solve for the condition that under the influence of the drive $\eps\hSig_1^x$, the two operators $\ket{\uparrow\downarrow}\bra{\downarrow\downarrow}$ and $\ket{\downarrow\uparrow}\bra{\downarrow\downarrow}$, which contribute to the $\ket{1}-\ket{3}$ and $\ket{1}-\ket{2}$ coherences, have vanishing expectation values. These conditions respectively lead to the following equations for $\gbar^*$:
\begin{align}
    &\ev{\ket{\uparrow\downarrow}\bra{\downarrow\downarrow}} = 0 \implies y_2-y_4 =0 \implies \gbar^*\left(\gbar^{*2} - \frac{3}{4} \right)=0,\nonumber\\
    &\ev{\ket{\downarrow\uparrow}\bra{\downarrow\downarrow}} = 0 \implies y_1-y_3 =0 \implies \gbar^{*2}\left(\gbar^{*2} - \frac{3}{4} \right)=0. 
\end{align}
Both the conditions are simultaneously satisified at $\gbar^*=0$ and $\gbar^*=\sqrt{3}/2$. The $\gbar^*=0$ point is trivial, since the two qubits are not interacting at this point. On the other hand, the point $\gbar^*=\sqrt{3}/2$ is non-trivial and the complete vanishing of coherences involving the state $\ket{\downarrow\downarrow}$ arises from an interplay of gain, loss and qubit-qubit interactions. Because of the symmetric choice of gain and loss parameters, the same solutions for $\gbar^*$ will be found if the drive $\eps\hSig_2^x$ is applied and we solve for the coherences involving the state $\ket{\uparrow\uparrow}$ to vanish.

\providecommand{\noopsort}[1]{}\providecommand{\singleletter}[1]{#1}%